\documentclass[10pt,twocolumn,twoside]{IEEEtran}

\usepackage[dvips]{graphicx}
\usepackage[fleqn]{amsmath}
\newcommand{\be}{\begin{equation}}
\newcommand{\ee}{\end{equation}}

\begin{document}

\title{Wavelet analysis: a new significance test for signals dominated by
intrinsic red-noise variability}

\author{Pawe{\l} Lachowicz,~\IEEEmembership{Member,~IEEE}
\thanks{P.~Lachowicz is with Centre for
Wavelets, Approximation and Information Processing at Temasek
Laboratories, National University of Singapore, 5A~Engineering Drive~1, \#09-02
Singapore~117411 (e-mail:pawel@ieee.org).}}

\markboth{Submitted to IEEE TRANSACTIONS ON SIGNAL PROCESSING}
{Lachowicz: A new significance test for wavelet analysis}

\maketitle

\begin{abstract}
We develop a new statistical test for the wavelet power spectrum.
We design it with purpose of testing signals which intrinsic variability
displays in a Fourier domain a red-noise component described by a single, broken
or doubly-broken power-law model. We formulate our methodology as
straightforwardly applicable to astronomical X-ray light curves and aimed at
judging the significance level for detected quasi-periodic oscillations (QPOs).
Our test is based on a comparison of wavelet coefficients derived for the source
signal with these obtained from the averaged wavelet decomposition of simulated
signal which preserves the same broad-band model of variability as displayed by
X-ray source. We perform a test for statistically significant QPO detection in
XTE~J1550--564 microquasar and active galaxy of RE~J1034+396 confirming these
results in the wavelet domain with our method. In addition, we argue on the
usefulness of our new algorithm for general class of signals displaying
$1/f^\alpha$-type variability.
\end{abstract}


\IEEEpeerreviewmaketitle

\section{Introduction}
\label{s:intro}

Fourier transform has become a very popular tool of time-series analysis within
last fifty years in almost all areas of research. Applied to test signal
frequency content, it was quickly recognized as a helpful method for
investigation of periodic variability \cite{bracewell78}. It gained a
particular interest in a wide field of astronomical research regarding different
classes of objects with variable emission. In particular, time-series provided
by recent high-energy detectors onboard X-ray orbiting NASA/ESA satellites
\cite{swank94}--\cite{xmm} have been found to trace remarkably rapid variability
coming from neutron star and black-hole systems \cite{ump}--\cite{rgc2000}.
Since the majority of X-ray emission takes its origin in accretion process of
matter and gas onto central object \cite{ss73}--\cite{urry95}, therefore the
observed signal variability is a direct manifestation of physical processes
taking place in a very strong gravitational field \cite{vdk04b}.

The accretion onto black-holes is one of the most energetic process in the
Universe \cite{fkr}. Energy spectra for the corresponding objects uncover both
thermal and non-thermal nature of emission (e.g. \cite{dg03}). The analysis of
Fourier power spectra (also referred to as Power Spectral Density; PSD) of X-ray
light curves $x(t)$ has revealed that the overall shapes are very often well
described, in general, by a power-law model, i.e. $P(f)\propto 1/f^\alpha$ where
$\alpha\in\Re$ denotes a slope (\cite{vdk04a}, \cite{pott03}--\cite{nowak99}).
In case of $\alpha> 0$ one claims about signals dominated by intrinsic
variability of red-noise type. The red-noise character of source variability is
generally rarely met in nature \cite{press78}. Nevertheless, it seems that
black-hole systems emitting in X-rays enlarge the sample significantly.

To date the majority of efforts devoted to our curiosity of understanding the
X-ray variability of astronomical objects were based on the systematic study of
PSD shapes and looking for occasionally observed quasi-periodic oscillations
(QPOs; \cite{vdk04a}, \cite{vdk05a}--\cite{vdk05b}). Since Fourier PSD contains
no information about time evolution of detected periodicities, the need of use
of the time-frequency techniques emerged. The application of
Short-Term Fourier Transform (STFT; e.g. \cite{cohen95}) provided a good
solution in this domain and has been found helpful in localizing QPOs (e.g.
\cite{wilms2001}--\cite{barret05}). However, regardless the frequency of
dectected QPOs, STFT keeps the same time-frequency resolution as dictated by a
length of a sliding window. In most cases it prevents a detection of
oscillations lasting shorter than the window span. On the contrary, a wavelet
analysis occurred to be more successful where the time-frequency resolution was
scale (frequency) dependant. Within recent fifteen years it has attracted
attention of many researchers \cite{mr93}--\cite{addison02}. 

The wavelet power spectrum (scalogram) for discrete evenly sampled X-ray signal
$x_n$ ($n=1,...,N$; $T\!=\!N\Delta t$) given as count-rate [cts~s$^{-1}$] can be
defined as the normalized square of the modulus of the wavelet transform:
\be
    W = \xi |w_n(a_m)|^2
\label{w}
\ee
where $\xi$ denotes a normalization factor and $w$ a discrete form of the
continuous wavelet transform in a function of two parameters: scale $a$ and
localized time index $n$ \cite{addison02}--\cite{tc98}:
\be
   w_n(a_m) = \left(\frac{2\pi a_m}{\Delta t} \right)^{1/2}
              \sum_{j=1}^{N}
              \hat{x}_j \hat{\psi}^\star(2\pi a_m f_j) e^{i2\pi f_j 
              n\Delta t}
\label{wkam}
\ee
where the discrete Fourier transform of signal $x_n$ is given by:
\be
   \hat{x}_j = \sum_{n=1}^{N}
               (x_n - \bar{x}) e^{-i(2\pi j n/N)}
\ee
and $j$ denotes a frequency index, 
\be
\nu_j =
\begin{cases}
\ \ \,  j/(N\Delta t) & \text{for $j\leq N/2$} \\
  -j/(N\Delta t) & \text{for $j>N/2$}
\end{cases}
\ee
and $\bar{x}$ a mean value of $x_n$ \cite{tc98}. Following \cite{vau03} one can
assume normalization factor of $w_n(a_m)$ to be $\xi=2\Delta t \bar{x}^{-2}$
which provides that integrated wavelet spectrum over time, also referred to as a
global wavelet power spectrum \cite{tc98},
\be
    G(a_m) = \frac{\xi}{N} \sum_{n=1}^{N} |w_n(a_m)|^2 ,
\ee
will hold the units of (rms/mean)$^2$ Hz$^{-1}$, i.e. it gives the variance,
relative to the mean, within a given frequency range of integration
\cite{vau03}--\cite{vdk89}. For the computation of (\ref{w}) the
use of a
dyadic grid of scales, i.e.
\be
    a_m = a_0 2^{m\Delta m}, \ \ \ \ m=0,...,M,
\ee
can be implemented \cite{tc98}, where the smallest scale $a_0=2\Delta t$
corresponds to the reversed Nyquist frequency, and
\be
    M = \Delta m^{-1} \log_2(N\Delta ta_0^{-1}).
\label{M}
\ee
A grid resolution is given by $\Delta m$. Since the search for periodicity and
quasi-periodicity in X-ray light curves seems to be the main target, it is
reasonable to assume the analyzing function to be Morlet wavelet, 
\be
\psi(t)=\pi^{-1/4} e^{i(2\pi f_0 t)} e^{-t^2/2},
\ee
which oscillates due to a term $\propto e^{{\rm i}t}$, where $2\pi f_0$
parameter is set to 6.

In order to popularize wavelet analysis as a new tool in research over
time-series, an additional requirement in the form of significance test was
strongly desired. Torrence \& Compo \cite{tc98} (TC98) were first who addressed
that issue extensively providing the public with an excellent guide and software
ready-to-use. In general, a map of wavelet power spectrum shows a distribution
of spectral components in the time--scale (or time--frequency) plane. Single
oscillation, of sufficiently high amplitude to be detected, marks itself in the
map as a wavelet peak. To every peak a certain statistical significance can be
assessed. We say that a wavelet peak is significant at assumed per cent of
confidence when it is above a certain background spectrum. The latter can be
defined by the mean Fourier power spectrum of analyzed time-series.

TC98 in their work adopted the background power spectrum, $P_j$, after
\cite{gilma63} (see also Eq.(16) in \cite{tc98}) which is valid for application
as long as the analyzed signal is considered to be a realization of univariate
lag-1 autoregressive (AR1) process \cite{marple}--\cite{kay}, $x_l = \alpha
x_{l-1} + z_l$, where $\alpha$ is lag-1 autocorrelation, $z_l$ denotes a random
variable drawn from the Gaussian distribution (with zero mean and variance
$\sigma^2$) and $x_0=0$. However, the PSD shape provided by \cite{gilma63} is
not a good representation of broad-band PSDs in case of considered here X-ray
sources. This is because of at least two reasons: (i) not every X-ray light
curve can be described by AR1 process \cite{umv05} (although \cite{katja98}),
(ii) PSDs are dominated by the red-noise spectral components which shapes are
mainly fitted with single, broken, doubly broken or even bending power-law
models.

TC98 showed that for a time-series modeled as the AR1 process its local wavelet
spectrum follows the mean Fourier spectrum given by their Eq.(16) \cite{tc98}.
By the ``local wavelet spectrum'' they define a vertical slice (along the
frequency axis) through a wavelet map. Empirical justification is done on the
way of Monte Carlo simulations which were performed considering only one local
wavelet power spectrum per simulation and, in addition, selected for a
particular moment of time ($k=256$ of 512 points). The corresponding
distribution of the local wavelet spectrum (at each localized time index $n$ and
scale $a_m$), ought to have $\chi^2_2$ distribution,
\be
     \frac{|w_k(a_m)|^2}{\sigma^2} \Rightarrow
     \frac{1}{2} P_j\chi^2_2, 
\ee
where $\chi^2_2$ denotes the $\chi^2$ distribution with two degrees of freedom
and an arrow ``$\Rightarrow$'' means ``is distributed as''. The above
implication is performed based on the assumption that a random variable $x_l$ is
normally distributed.

A noteworthy work in this domain has been recently done by Zhongfu Ge
\cite{ge07} who generalized TC98's significance test. Ge left the assumption of
signal to be the realization of AR1 process and proposed Gaussian White Noise
process to serve as the null hypothesis.

What TC98 and Ge did not show is that the local wavelet power spectra, $P_j$, at
nearby times $n$ are correlated due to variable length of the wavelet function,
and the correlation is stronger at low frequencies as shown by \cite{mk04}. This
introduces departures off $\chi^2$ distribution in any finite data set. Thus a
more correct procedure is required to take this effect into account.

Maraun, Kurths \& Holschneider \cite{mkh07} proposed another test for wavelet
significance, namely, {\it areawise test} which aims at comparing the wavelet
peak size to the expected  time-frequency uncertainty. They noticed that for any
process with unknown distribution a number of wavelet peaks forming in the
wavelet map different shapes (circles, ovals, patches, etc.) become difficult in
interpretation. The extracted picture is messy. Some peaks are due to red-noise
process of the source intrinsic variability and the others are due to
white-noise. As shown by them, the areawise test may reduce a number of
significant peaks even by 90 per cent.

To judge the reality of each wavelet peak it seems best to set the significance
level highly enough in order to reject spurious results. However, not always
this is a right way, for instance, in case of X-ray sources of rather low
count-rate (low S/N ratio) and strong mixing between red- and white-noise
components in high-frequency region. On the other hand, this also does not mean
that the analysis does not contain any scientific results. For example, the same
situation took place for COBE detections of microwave background peaks
\cite{smoot}. Most of them were spurious but the fluctuations were statistically
different from expectations of pure noise \cite{fabbri96}.

None of this method can be applied as a fully trustful test working well for
astronomical X-ray signals as long as their assumptions rely on the analyses
that use impropriete background spectra. In this paper we address that issue
and propose a new significance test.

\section{Wavelet significance for red-noise signals \\ of power-law form}
\label{s:xsig}

We design a method for determination of significance of wavelet spectral
features (in short, wavelet peaks) calculated for any X-ray light curve which
intrinsic variability is a realization of a sum of red- and white-noise
(Poisson) processes. The main idea standing for that is to compare the
distribution of wavelet power between real data and simulated signal.
Here, by the simulated light curve (also referred to as TK light curve) we will
understand a time-series which displays the same profile of red-noise
variability as that one found for X-ray data. Before providing the reader with
ready-to-use algorithm (Sect.~\ref{ss:algorithm}), some essential information on
the background assumptions need to be clarified. We provide them within the
following two subsections (Sect.~\ref{ss:simu} and \ref{ss:best}).

\subsection{Simulations of $1/f^\alpha$ process}
\label{ss:simu}

Let's assume that $x(t)$ describes a part of stochastic continuous process
$X(t)$ generated by the physical system of unknown nature which produces the
variability. Based solely on $x(t)$ a reproduction of the overall information
and properties of the system is very difficult. Recorded data can facilitate a
determination of a model which would reproduce the system best. The observed
light curve of X-ray emission is the realization of the physical process(es) for
which calculated Fourier power spectrum (PSD) reveals the variability of
power-law $1/f^\alpha$-form (see references in Section~\ref{s:intro}).
Therefore, we are able to approximate one of the system's parameters (i.e. its
variability) by a power-law model which, as we believe, constitutes the best PSD
description of the underlying process.

Aperiodic broad-band variability as displayed in X-ray light curves can be
thought, in the first approximation, as the realization of a linear stochastic
process, weakly non-stationary. Timmer \& Konig \cite{tk} proposed an algorithm
(hereafter also referred to as TK) for generating linear aperiodic signals,
$l(t)$, that exhibit $P(f) \propto 1/f^\alpha$ power-law spectrum,
\be
	l(t) \propto \sum_f \sqrt{P(f)} \cos\left[2\pi f - \phi(f)\right] ,
\label{tkx}
\ee
where the phase is randomized $\phi(f)\in [0,2\pi]$. Since the amplitudes
are taken as $\sqrt{P(f)}$, \cite{tk} noticed that in order to create power-law
signal it is essential to randomize also the amplitude. 

A light curve obtained in this manner displays $1/f^\alpha$-shape, however as
argued by \cite{umv05}, it does not preserve fundamental statistical properties
of X-ray time-series for accreting black-hole systems. The latter have been
found to be rather non-linear showing rms--flux relation at all time-scales
\cite{umv05}, \cite{um01} (rms is defined as a square root of variance 
calculated for a signal). According to \cite{umv05}, formally non-linear type
of variability for any linear input light curve can be obtained by taking simple
exponential transformation of (\ref{tkx}), i.e.~$x_{\rm exp}(t)=\exp[l(t)]$.  To
make the picture complete every simulated light curve should include white
noise. Following \cite{lc05} it can be expressed as:
\be
	x(t) = \mbox{\sc poisdev}[x_{\rm exp}(t)\Delta t]/\Delta t 
\ee
where $x(t)$ denotes simulated signal with Poisson noise, $\Delta t$ is
a light curve resolution (bin time) and {\sc poisdev} represents a subroutine
generating a random number with Poisson distribution \cite{press}.

\subsection{Selection of best $1/f^\alpha$ model for PSD of X-ray source}
\label{ss:best}

A description of Fourier PSD of X-ray object under investigation can be obtained
in the process of numerical fitting of power-law model. As reviewed by
\cite{vdk04a}, a small but substantial subset of X-ray black-hole (and neutron
star) systems apart from exhibiting  $1/f^\alpha$-like broad-band variability,
also show quasi-periodic oscillations (QPOs). Hardly ever met as
tiny spikes, they are much often resolved as broadened features characterized by
the quality factor of $Q=f/\Delta f$ where $\Delta f$ denotes the peak's full
width at half maximum (FWHM) and $f$ QPO central frequency. The broader peak
the higher $\Delta f$ thus the lower $Q$. So far, the majority of QPO peaks have
been found to be surprisingly well described by Lorentzians, e.g. \cite{pott03},
\cite{nowak99}. Lorentzians are defined as $L(f)\propto
\Delta f/[f-f_0)^2 +(\Delta f/2)^2]$ with a centroid frequency $f_0$ and, if
detected, they constitute an intergal component of the overall PSD model.
In practice, fitting of power-law model as a background model may or may not
include fitting of $L(f)$ component simultaneously. In the latter case, PSD data
points describing Lorentzian can be simply omitted and these two aforementioned
approaches rely on the individual preferences and fitting experience.

In this work our wish will be to generate simulated X-ray light curve which
will own the same power-law background model of variability as fitted to
the object's PSD. This step requires that the overall rms and the average count
rate for simulated time-series must match the corresponding same values for
X-ray signal. For Fourier PSD normalized to unit of (rms/mean)$^2$ Hz$^{-1}$
\cite{vau03}, the root-mean-square (rms) can be obtained by integration of
underlying PSD in a given band of frequencies
\cite{vdk89} as follows:
\be
    \mbox{rms}^2 = \sigma^2 = \int_{f_1}^{f_2} P(f) df \approx
		     \sum_{j_1}^{j_2} P_j \Delta f
\label{rms}
\ee
where $1/N\Delta t \le f_1 < f_2 \le 1/2\Delta t$. We argue that rms values,
both obtained for simulated signal and X-ray light curve ought to be calculated
in the same $(f_1,f_2)$ range of frequencies what can guarantee the best match.
The integration (\ref{rms}) should omit PSD data points corresponding to QPO
component(s), if present.

\subsection{A new algorithm for wavelet significance}
\label{ss:algorithm}

We define a new algorithm leading to determination of significance matrix
for derived wavelet power spectrum by the following steps:

\begin{itemize}
\item
Two-dimensional matrix of wavelet power spectrum (\ref{w}) for
a time-series $x_n$ is a set of $(M+1)\times N$ wavelet coefficients where $M$
corresponds to (\ref{M}) and $N$ denotes a number of data points. Let us denote
the results of computation of (\ref{w}) for X-ray source as:
\be
   \ \ \ \ W^{\rm src}_{(M+1)\times N} \ .
\ee
Please notice that in practice instead of time--scale plane one often refers to
wavelet spectrum displayed in time--frequency plane. If required, for Morlet
wavelet, a transformation from scales to Fourier frequencies can be applied
according to $f_j=(1.03a_m)^{-1}$ relation \cite{tc98}.
\item
For a given discrete signal $x_n$ calculate its Fourier power spectrum and find
the best model which describes it (Section~\ref{ss:best}). If the model for
considered data set is known in advance (e.g. from the existing literature), use
the best fitted parameters as the reference ones.
\item
Generate one long simulated light curve (with the same $\Delta t$), best at
least two or three orders of magnitude longer than your signal under analysis.
Use Timmer \& K\"{o}nig's method which allows to obtain simulated signals
from Fourier power spectra of assumed shape (Section~\ref{ss:simu}). Take care
to keep the rms variability at the same level as in the data
(Section~\ref{ss:best}).
\item
From the simulated TK light curve, select randomly $K$ signals of the same time
duration as in original data set, i.e. $N\Delta t$, and compute for
each of them the wavelet power spectrum according to (\ref{w}). Shall us
denote the resulting matrices as:
\be
    \ \ \ \ W^i_{(M+1)\times N} \ \ \mbox{where} \ \ i=1,...,K.
\label{Wi}
\ee
In calculation of (\ref{Wi}) for normalization factor $\xi$ (see
Sect.~\ref{s:intro}) assume every time a mean count rate $\bar{x}$ of the 
local TK signal $i$.
\item
Find the maximum value between all $W^i$ matrices:
\be
   \forall \ W^i \ \exists \ x \ \ \mbox{where} 
   \ \ x=\max(W^i) .
\ee
\item
For every scale $m$ of every $W^i$ calculate the histogram of wavelet
power distribution from 0 to $x$ with a histogram resolution defined as $\Delta
h=x/(M+1)$. In consequence, the following matrices shall be created:
\be
   \ \ \ \ H^i_{(M+1)\times(x/\Delta h)}
\ee
consisting of $M+1$ histograms.
\item
Calculate a matrix containing averaged histograms for every scale $m$:
\be
   \ \ \ \ \bar{H}_{(M+1)\times(x/\Delta h)} = \frac{1}{K}
   \sum_{i=1}^{K} H^i.
\ee
This step is introduced in order to smooth out different distribution of
wavelet power appearing in every $H^i$ for each scale. Please note the higher
$K$ the more smoothed averaged wavelet power histograms one obtains.
\item
For every of $M+1$ histograms stored in $\bar{H}$, determine the $p$-th
quantile $x_p$. Namely, let $\bar{H}(m)$ be a continuous random variable, then
for $0<p<1$, the $p$-th quantile is the $x_p$ such that $\mbox{Pr}(\bar{H}(m)\le
x_p) = p$ \cite{clapham96}. For instance, in deriving of the significance
of wavelet peaks at the 95 per cent confidence level one aims in determination
of 0.95-th quantile $x_p$ in a function of scale $m$ (frequency). In practice, a
numerical integration of normalized\footnote{A term {\it normalized} means that
an integral of $\bar{H}(m)$ must equal 1.} $\bar{H}(m)$ must be
performed until an integral value exceeds 0.95. This determines quantile $x_p$
with an error of $\Delta h$ for each
histogram:
\be
    \ \ \ \
    X_p = \left[
          \begin{array}{c}
            x_p^0 \\
            x_p^1 \\
            ...   \\
            x_p^m \\
            ...   \\
            x_p^{M}
          \end{array}
          \right]_{(M+1)\times 1} \ .
\label{xp} 
\ee
\item
Expand matrix $X_p$ to have dimensions $(M+1)\times N$ by
calculating a linear algebraic product of the matrices $X_p$ and $I$:
\be
   \ \ \ \ W^{\rm bkg}_{(M+1)\times N_{\rm obs}}(i,j) =
           \sum_{k=1}^{1} X_p(i,k) I(k,j) ,
\ee
where $I$ is a matrix of ones of $1\times N$ dimension. Denote 
the result as a background wavelet matrix $W^{\rm bkg}$.
\item
Finally, a matrix of $(1-p)100\%$ significance, $S$, will identify
significant wavelet peaks appearing in computed wavelet power spectrum $W^{\rm
src}$ when the following condition will be met:
\be
   \ \ \ \ S_{(M+1)\times N_{\rm obs}} = \frac{W^{\rm src}}{W^{\rm bkg}} > 1.
\label{smatrix}
\ee
\end{itemize}

In proposed procedure a distribution of wavelet power at a given wavelet scale
(frequency) is directly compared to the distribution of power obtained for
simulated light curves. Since the Fourier PSD model and rms variability in
simulated time-series are assumed to match the same quantities as for the X-ray
source data, they serve as a good reference level to test any excess of wavelet
power above it. In addition, by testing a wavelet power scale by scale, one
takes into account an influence of correlation of wavelet power over a range of
frequency and time due to varying length of a wavelet function
(Sect.~\ref{s:intro}, \cite{mk04}).

\section{Application to X-ray observations}

In this section we provide the reader with two examples of application of our
new algorithm to real observations of black-hole systems emitting in X-rays:
microquasar XTE~J1550--564 \cite{sm98} and active galactic nuclei of
RE~J1034+396. In case of the former object, the source is known because of
displaying a strong X-ray aperiodic variability with a variety of low-frequency
(0.08--18~Hz) and high-frequency QPOs (100--285~Hz) \cite{orosz}--\cite{rem99}.
Here, we examine its activity peaked at the frequency of 4~Hz and derive
corresponding QPO significance in the wavelet domain. As for the latter
source, we select the active galaxy of RE~J1034+396 which has been recently
found to uncover the very first ever significant signature (over $3\sigma$) of
quasi-periodicity among all active galaxies \cite{gier08}. Inspired by Fourier
results of \cite{gier08}, we verify their findings by determination of wavelet
significance for detected $\simeq 2.7\times 10^{-4}$~Hz variability.

\subsection{XTE~J1550--564}
\label{ss:j1550}

In order to examine XTE~J1550--564 light curve variability which
has been dominated by intrinsic red-noise component in
$10^{-2}$--$10^{1}$~Hz frequency band, after \cite{zg04} we select {\it
RXTE}/PCA observation of 40099-01-24-01 (1998-09-29) from the public HEASARC
archive. We reduce the data with LHEASOFT package ver.~6.6.1 applying the
standard PCA selection criteria for above data set (see
http://heasarc.gsfc.nasa.gov for details). As the final product we extract a
2--13~keV light curve with a bin size of $\Delta t=2^{-5}$~s for the first part
of PCA observation (i.e. $N=80065, T=N\Delta t= 2502$~s).

\begin{figure}
\includegraphics[angle=-90,width=0.45\textwidth]{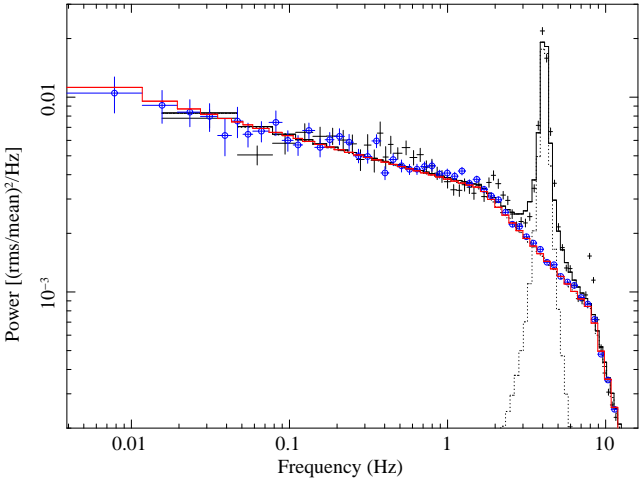}
\vspace*{30pt}
\caption{Power spectrum density of XTE~J1550--564 as observed by {\it RXTE}/PCA
on 1998-09-29. Black solid line denotes fitted model
composed of double broken power-law model (red solid line) and Lorentzian
(dotted line) representing red-noise background spectrum and
quasi-periodic oscillation (QPO), respectively. Power spectrum for simulated TK
time-series has been marked in blue (circle markers).}
\label{fig:psdj1550}
\end{figure}

Fig.~\ref{fig:psdj1550} presents Fourier power spectrum density calculated
from 78 averaged spectra based on 1024 point data segments. The overall PSD
shape can be split into two main components: power-law underlying spectrum and
broad line profile representing red-noise source variability and QPO component,
respectively. In the $\chi^2$ fitting process we found that a model composed of
three-segment broken power-law (i.e. with two break frequencies) and Lorentzian
profile line \cite{arnaud} fit the PSD best yielding reduced-$\chi^2$ of 10.9 at
59 degrees of freedom. The corresponding power-law slope values and break
frequencies have been determined as follows: $\alpha_1=0.22\pm 0.02,
\alpha_2=0.90\pm 0.02, \alpha_3=3.40\pm 0.07, f_1=1.63\pm 0.08$~Hz and
$f_2=8.12\pm 0.10$, whereas for Lorentzian the central frequency of
$f_0=4.104\pm 0.006$~Hz and its FWHM $\Delta f=0.33\pm 0.01$~Hz were fitted
providing QPO quality factor of $Q\simeq 12.4$. We marked both model components
in Fig.~\ref{fig:psdj1550} by red solid line and dotted line, respectively,
where the overall best model has been indicated by black solid line.

Based on the above power-law model we generated a $2^{16}$ point long simulated
signal (64 segments of 1024 points) with TK method. In simulation we assumed the
sampling time of $\Delta t=2^{-5}$~s and ensured that resulting rms value (here,
derived in 0.5--2~Hz and 8--12~Hz frequency band; see Sect.~\ref{ss:best} for
details) was nearly the same as compared to corresponding rms values for X-ray
source. We calculated the averaged Fourier PSD for TK time-series and plotted it
in Fig.~\ref{fig:psdj1550} by blue circle markers. One can clearly notice that
TK PSD reproduces well the assumed double broken power-law model component.

\begin{figure}
\includegraphics[angle=0,width=0.475\textwidth]{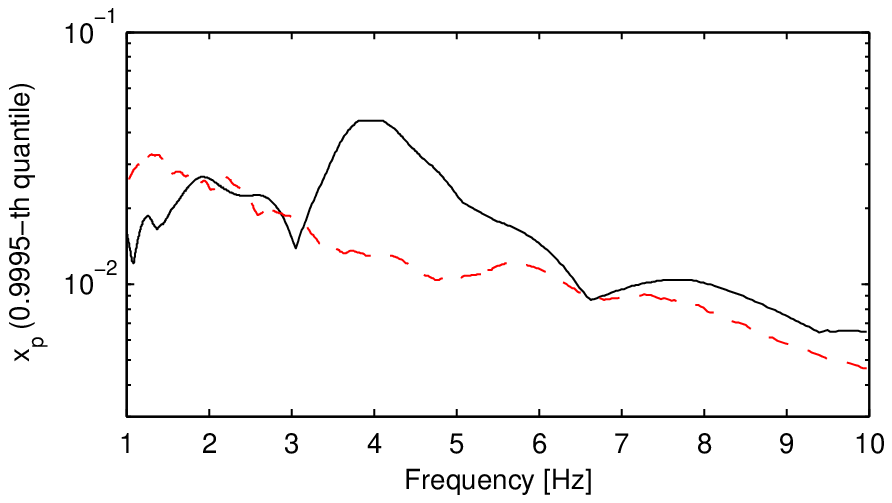}
\includegraphics[angle=0,width=0.475\textwidth]{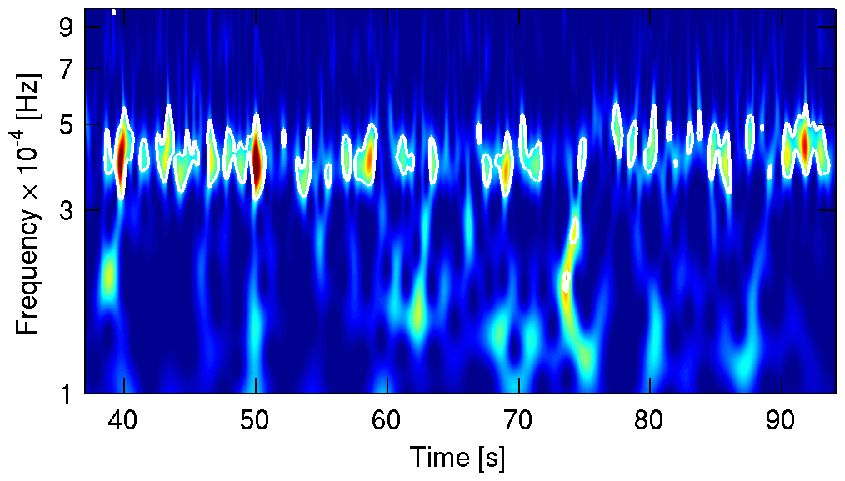}
\caption{{\it (upper)} The dependence of 0.9995-th quantiles in a
function of frequency calculated for selected 96~s data signal of
XTE~J1550--564 (solid black line) and obtained from wavelet analysis of
simulated TK series (red dashed line). See Section~\ref{ss:j1550} for details.
{\it (bottom)} Corresponding wavelet power spectrum displayed with derived
significance contours (0.9995 confidence level). For better clarity of
map reading we plot 96~s long wavelet spectrum between 37--94~s only.}
\label{fig:waj}
\end{figure}

Following the recipe given in Section \ref{ss:algorithm} we aim at
determination of wavelet confidence at 0.9995 level. From the entire
XTE~J1550--564 light
curve, we select a 3072 point long {\it test} signal (96~s) for which we compute
wavelet significance matrix (Eq.~\ref{smatrix}). Upper plot of
Fig.~\ref{fig:waj} presents a dependence of 0.9995-th wavelet quantile
(Eq.~\ref{xp}) in a function of Fourier frequency in 1--10~Hz band, both for
X-ray data segment (black solid line) and TK simulation (red dashed line). It
is straightforward to note that at the significance of $5\times 10^{-4}$
the quasi-periodic variability around $f\simeq 4$~Hz is high
above the mean (simulated) red-noise level. Finally, the bottom map of
Fig.~\ref{fig:waj} presents the wavelet power spectrum (Eq.~\ref{w})
for considered data segment with overlaid contours of derived significance.

\subsection{RE~J1034+396}

\begin{figure}
\includegraphics[angle=0,width=0.485\textwidth]{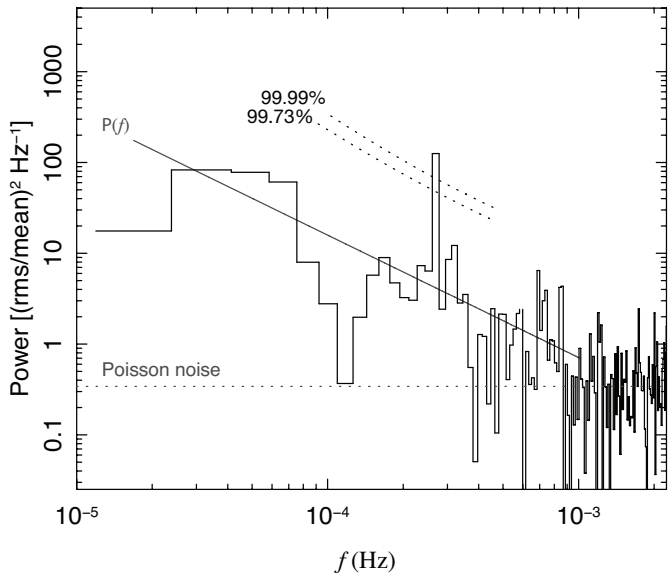}
\caption{Fourier power spectral density of an active galaxy of RE~J1034+396
\cite{gier08} displaying a significant quasi-periodicity at frequency $f\simeq
2.7\times 10^{-4}$~Hz $(>99.99\%$ of confidence; upper dotted line). Solid
line denotes best power-law model fitted to the data $P(f)\propto f^{-1.35}$,
whereas bottom horizontal dotted lines marks expected white-noise level.
Reproduced with kind permission of Marek Gierlinski.}
\label{fig:rej}
\end{figure}

A Seyfert~1 galaxy of RE~J1034+396 has been found and confirmed as the
first active galaxy showing highly significant quasi-periodic
oscillation (QPO) at the frequency $f\simeq 2.7\times 10^{-4}$~Hz
\cite{gier08}. We reproduce their PSD in Fig.~\ref{fig:rej} as computed for
signal segment of {\it XMM-Newton} EPIC observation of RE~J1034+396 conducted on
2007-05-31. Solid black line crossing PSD denotes best-fitted power-law spectrum
described as:
\be
    \lg P(f)=-1.35\lg f-4.128 
\label{lgP}
\ee
(M. Gierlinski, private communication). Upper two dashed lines mark computed
confidence levels for detected periodicity at 99.73\% ($3\sigma$) and 99.99\%
confidence levels. In order to verify aforementioned significance of QPO
detection, we downloaded from {\it XMM-Newton} Science Operations Centre the
same EPIC pn+MOS1/MOS2 observation data set and reduced it in the same manner as
described in \cite{gier08}. We limit original X-ray light curve (0.3--10~keV)
to 601 point long time-series (corresponding to segment 2 of \cite{gier08})
where the sampling time of $\Delta t=100$~s has been used.

\begin{figure}
\vspace*{-15pt}
\includegraphics[angle=0,width=0.485\textwidth]{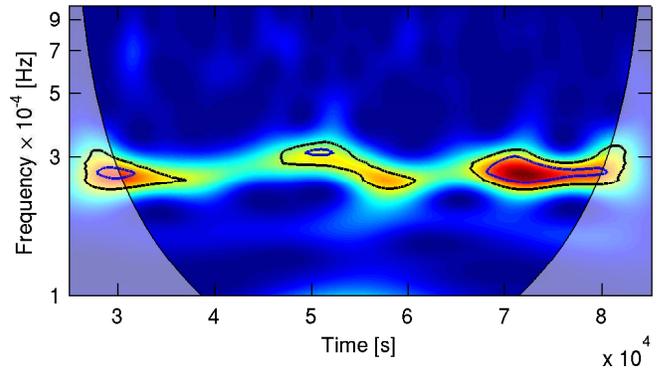}
\caption{RE~J1034+396: Wavelet power spectrum calculated for the second data
segment of {\it XMM-Newton} EPIC/pn+MOS observation, as defined by
\cite{gier08}. Quasi-periodic modulation is clearly detected at $f\simeq
2.7\times 10^{-4}$~Hz frequency and last throughout a whole light curve
duration. Outer and inner contours correspond to derived 99.73\%
($3\sigma$) and 99.99\% confidence level, respectively, with the use of new
algorithm as provided in Section~\ref{ss:algorithm}.}
\label{fig:warej}
\end{figure}

We compute wavelet power spectrum (\ref{w}) in $10^{-4}$--$10^{-3}$~Hz band and
make use of our new algorithm in order to determine two contour plots marking
99.73\% and 99.99\% confidence for wavelet results.
For the purpose of wavelet background simulation, we generate $2^{16}$ point TK
signal based on assumed single power-law model (Eq.~\ref{lgP}). Since the
underlying PSD model is known in advance, we perform an integration of it
between $10^{-4}$~Hz and $10^{-3}$~Hz to find the corresponding rms value in
this frequency range (rms$\simeq 5.5$\%). Within the simulation process
of TK time-series we check to make sure that in the same frequency band the rms
of TK signal stays in agreement with derived value. Based on that, we determine
two significance matrices (Eq.~\ref{smatrix}) and plot them together with
wavelet power spectrum. The final form of all calculations presents
Fig.~\ref{fig:warej} where an outter and inner contour denote 99.73\% and
99.99\% confidence level, respectively.

Remarkably, a detected quasi-periodic oscillation at $f\simeq 2.7\times
10^{-4}$~Hz appears to be very coherent in time though showing small but
noticeable period drift. We checked that the entire periodicity remains
significant at the confidence level of 99\% throughout the signal segment
duration. In contrast, as marked in Fig.~\ref{fig:warej}, only a small
fraction of QPO activity exceeds top level of $0.01\%$ significance whereas
about 70\% of detected QPO signal stays significant at $3\sigma$ level.

\section{Conclusions}

In this paper we reviewed a recent status of knowledge devoted to wavelet
significance tests and proposed a new method designed especially for signals
which intrinsic variability displays in Fourier power spectral density a
power-law form. Since a large number of astronomical X-ray sources has been
found to be characterized by the similar PSD shapes, we focused our attention to
that class of objects and their time-series analysis.

We presented a flexible algorithm helpful in determination of the significance
of quasi-periodic oscillations in the wavelet domain. Our method has been based
on a comparison of wavelet coefficients calculated for X-ray time-series
with those computed for simulated signal, where the latter displays the same
PSD broad-band model of variability as X-ray object.

We argue that the usefulness of our new algorithm can be easily extended to
non-astronomical time-series which PSD can be described by power-law model as
well. If required, the omission of exponential transformation of linear signal
generated with the use of TK method and inclusion of Poisson noise should
be avoided (Section~\ref{ss:simu}).

\section*{Acknowledgment}

The author would like to thank Prof. Didier Barret for inspiration which led
to the accomplishment of this paper and Dr. Marek Gierli{\'n}ski for a kind
providing with an original version of Fig.~\ref{fig:rej}.


\begin{thebibliography}{3}

\bibitem{bracewell78} 
R. N. Bracewell, "The Fourier Transform and Its Applications", McGraw-Hill,
1965, 2nd ed. 1978, revised 1986

\bibitem{swank94} 
J. H. Swank, ``The XTE Mission'', {\it Bulletin of the American Astronomical
Society}, vol. 26, pp. 1420, 1994 

\bibitem{xmm} 
C. Gabriel, M. Guainazzi and L. Metcalfe, ``XMM-Newton: Passing Five Years
of Successful Science Operations'', Astronomical Data Analysis Software
and Systems XIV, vol. 347, pp. 425, 2005

\bibitem{ump} 
P. Uttley, I. M. McHardy and I. E. Papadakis, ``Measuring the broad-band
power spectra of active galactic nuclei with RXTE'',
{\it Monthly Notices of the Royal Astronomical Society}, vol. 332, issue 1, pp.
231-250, 2002

\bibitem{vdk04a} 
W. H. G. Lewin and M. van der Klis, ``Compact stellar X-ray sources'', Cambridge
Astrophysics Series, No. 39. Cambridge, UK: Cambridge University Press, 2006

\bibitem{rgc2000}
R. Sunyaev, and M. Revnivtsev, ``Fourier power spectra at high frequencies: a
way to distinguish a neutron star from a black hole'', {\it Astronomy and
Astrophysics}, vol. 358, pp. 617-623, 2000

\bibitem{ss73} 
N. I. Shakura and R.A. Sunyaev, ``Black holes in binary
systems. Observational appearance'', {\it Astronomy and Astrophysics}, vol. 24,
pp. 337, 1973

\bibitem{urry95}
C. M. Urry and P. Padovani, ``Unified Schemes for Radio-Loud Active Galactic
Nuclei'', {\it Publications of the Astronomical Society of the Pacific}, vol.
107, p. 803

\bibitem{vdk04b} 
M. van der Klis, ``Neutron Star QPOs as Probes of Strong Gravity and Dense
Matter'', X-ray Timing 2003: Rossie and Beyond. AIP Conference
Proceedings, vol. 714, pp. 371-378, 2004

\bibitem{fkr}
J. Frank, A. King A. and D.~J. Raine, ``Accretion Power in Astrophysics'',
Cambridge University Press, 2002

\bibitem{dg03} 
C. Done and M. Gierli{\'n}ski, ``Observing the effects of the
event horizon in black holes'', {\it Monthly Notice of the Royal Astronomical
Society}, vol. 342, issue 4, pp. 1041-1055, 2003

\bibitem{pott03}
K. Pottschmidt, J. Wilms, M. A. Nowak, G. G. Pooley, T. Gleissner, W. A. Heindl,
D. M. Smith, R. Remillard and R. Staubert, ``Long term variability of Cygnus
X-1. I. X-ray spectral-temporal correlations in the hard state'', {\it Astronomy
and Astrophysics}, vol. 407, pp. 1039-1058, 2003

\bibitem{law87} 
A. Lawrence, M. G. Watson, K. A. Pounds,and M. Elvis, ``Low-frequency
divergent X-ray variability in the Seyfert galaxy NGC4051", {\it Nature},
vol. 325, p. 694, 1987

\bibitem{mch04}
I. M. McHardy, I. E. Papadakis, P. Uttley, M. J. Page and K. O. Mason,
``Combined long and short time-scale X-ray variability of NGC 4051 with RXTE and
XMM-Newton'', {\it Monthly Notices of the Royal Astronomical Society}, vol. 348,
issue 3, pp. 783-801, 2004

\bibitem{bh90} 
T. Belloni and G. Hasinger, ``Variability in the noise properties of Cygnus
X-1'', {\it Astronomy and Astrophysics}, vol. 227, no. 2, L33-L36, 1990

\bibitem{nowak99}
M. A. Nowak, B. A. Vaughan, J. Wilms, J. B. Dove and M. C. Begelman,
``Rossi X-Ray Timing Explorer Observation of Cygnus X-1. II. Timing Analysis'',
{\it The Astrophysical Journal}, vol. 510, issue 2, pp. 874-891, 1999

\bibitem{press78} 
W. Press, "Flicker noises in astronomy and
elsewhere", Comments on Astrophysics, vol. 7, p. 103, 1978

\bibitem{vdk05a} 
M. van der Klis, ``The QPO phenomenon'', {\it Astronomische Nachrichten},
vol. 326, issue 9, p.798-803, 2005

\bibitem{vdk05b} 
M. van der Klis, ``Comparing Black Hole and Neutron Star Variability'', {\it
Astrophysics and Space Science}, vol. 300, issue 1-3, pp. 149-157, 2005

\bibitem{cohen95}
L. Cohen, ``Time-frequency analysis'', Englewood Cliffs, NJ: Prentice-Hall, 1995

\bibitem{wilms2001}
J. Wilms, M. A. Nowak, K. Pottschmidt, W. A. Heindl, J. B. Dove and M. C.
Begelman, ``Discovery of recurring soft-to-hard state transitions in LMC X-3'',
{\it Monthly Notices of the Royal Astronomical Society}, vol. 320, issue 3, pp.
327-340, 2001 

\bibitem{barret05}
D. Barret, W. Klu{\'z}niak, J.~F. Olive, S. Paltani and G.~K. Skinner, ``On the
high coherence of kHz quasi-periodic oscillations'', {\it Monthly Notices of the
Royal Astronomical Society}, vol. 357, issue 4, pp. 1288-1294, 2005

\bibitem{mr93} 
Y. Meyer and S. Roques, ``Progress in wavelet analysis and
applications'', 1993, Proceedings of the International Conference ''Wavelets and
Applications'', Toulouse, France, June 1992, Gif-sur-Yvette: Editions
Frontieres, ed. by Meyer, Yves; Roques, Sylvie, 1993

\bibitem{addison02}
P. S. Addison, ``Illustrated Wavelet Transform Handbook'', IoP, 2002

\bibitem{lc05}
P. Lachowicz and B. Czerny, ``Wavelet analysis of millisecond variability
of Cygnus X-1 during its failed state transition'', {\it Monthly Notices of the
Royal Astronomical Society}, vol. 361, issue 2, pp. 645-658, 2005

\bibitem{tc98} 
C. Torrence and G. P. Compo, ``A Practical Guide to Wavelet
Analysis'', {\it Bulletin of the American Meteorological Society}, vol. 79,
issue 1, pp. 61-78, 1998

\bibitem{vau03}
S. Vaughan, R. Edelson, R. S. Warwick and P. Uttley, ``On characterizing the
variability properties of X-ray light curves from active galaxies'', {\it
Monthly Notices of the Royal Astronomical Society}, vol. 345, issue 4, pp.
1271-1284, 2003

\bibitem{vdk89}
M. van der Klis, ``Quasi-periodic oscillations and noise in low-mass X-ray
binaries'', {\it Annual review of astronomy and astrophysics}. Volume 27. Palo
Alto, CA, Annual Reviews, Inc., pp. 517-553, 1989

\bibitem{gilma63}
D. L. Gilman, F. J. Fuglister and J. M. Mitchell Jr., ``On the Power
Spectrum of Red Noise'', {\it Journal of Atmospheric Sciences}, vol. 20, issue
2, pp. 182-184, 1963

\bibitem{marple}
S. L. Marple, Jr., ``Digital Spectral Analysis with Applications'', Prentice
Hall, Englewood Cliffs, 1987, Chapter 8

\bibitem{chatfield}
C.~Chatfield, ``The Analysis of Time Series. An Introduction.'', Chapman \&
Hall, A CRC Press Company, Sixth Ed., 2004

\bibitem{kay}
S. M. Kay, ``Modern Spectral Estimation: Theory and Application''. Englewood
Cliffs, NJ: Prentice-Hall, 1988

\bibitem{umv05} 
P. Uttley, I. M. McHardy and S. Vaughan, ``Non-linear X-ray variability
in X-ray binaries and active galaxies'', {\it Monthly Notices of the Royal
Astronomical Society}, vol. 359, p. 345, 2005

\bibitem{katja98} 
K. Pottschmidt, M. Koenig, J. Wilms and R. Staubert, ``Analyzing short-term
X-ray variability of Cygnus X-1 with Linear State Space Models'',
{\it Astronomy and Astrophysics}, vol. 334, p. 201, 1998

\bibitem{ge07} 
Z. Ge, ``Significance test for wavelet power and the
wavelet power spectrum'', {\it Ann. Geophys.} , vol. 25, p. 2259, 2007

\bibitem{mk04}
D. Maraun and J. Kurths, ``Cross Wavelet Analysis. Significance Testing and
Pitfalls'', {\it Nonlin. Proc. Geoph.}, vol. 11(4), pp. 505-514, 2004

\bibitem{mkh07}
D. Maraun, J. Kurths and M. Holschneider, ``Nonstationary Gaussian processes
in wavelet domain: Synthesis, estimation, and significance testing'', {\it
Physical Review E}, vol. 75, issue 1, id. 016707, 2007
 
\bibitem{smoot} 
G.~F. Smoot, et al., ``Structure in the COBE differential
microwave radiometer first-year maps'', {\it Astrophysical Journal},
vol. 396, L1 , 1992

\bibitem{fabbri96}
R. Fabbri and S. Torres, ``Peak statistics on COBE maps'', {\it Astronomy and
Astrophysics}, vol. 307, pp. 703-707, 1996

\bibitem{tk}
J. Timmer and M. Koenig, ``On generating power law noise'', {\it Astronomy and
Astrophysics}, vol. 300, p. 707, 1995

\bibitem{um01} 
P. Uttley and I. M. McHardy, ``The flux-dependent amplitude of
broadband noise variability in X-ray binaries and active galaxies'',
{\it Monthly Notices of the Royal Astronomical Society}, vol. 323, L26, 2001

\bibitem{press}
W. H. Press, S. A. Teukolsky, W. T. Vetterling W.T. and B. P. Flannery B.P.,
``Numerical recipes in FORTRAN. The art of scientific computing'', Cambridge:
University Press, 2nd ed., 1992

\bibitem{vdk97} 
M. van der Klis, ``Quantifying Rapid Variability in Accreting Compact Objects'',
Statistical Challenges in Modern Astronomy II, 321, 1997

\bibitem{clapham96}
Ch. Clapham, The Concice Oxford Dictonary of Mathematics, Oxford
University Press, 1996

\bibitem{sm98} 
D.~A. Smith, ``XTE J1550-564'', IAU Circ., 7008, 1, 1998

\bibitem{orosz} 
J.~A. Orosz, et al.\, ``Dynamical Evidence for a Black Hole in
the Microquasar XTE J1550-564'', {\it The Astrophysical Journal}, vol. 568,
p. 845, 2002

\bibitem{cui99} 
W. Cui, S. N. Zhang, W. Chen and E.~H. Morgan, ``Strong
aperiodic X-ray variability and quasi-periodic oscillationin X-ray nova XTE
J1550-564, {\it The Astrophysical Journal Letters}, vol. 512, L43, 1999

\bibitem{rem99} 
R.~A. Remillard, J.~E. McClintock, G.~J. Sobczak, C. D. Bailyn, J. A. Orosz,
E.~H. Morgan and A.~M. Levine, ``X-Ray Nova XTE J1550-564: Discovery of a
Quasi-periodic Oscillation near 185 Hz'', {\it The Astrophysical Journal
Letters}, vol. 517, L127, 1999

\bibitem{gier08} 
M. Gierli{\'n}ski, M. Middleton, M. Ward, and C. Done,
``A periodicity of $\sim$1hour in X-ray emission from the active galaxy RE
J1034+396'', {\it Nature}, vol. 455, issue 7211, pp. 369-371, 2008

\bibitem{zg04} 
A. A. Zdziarski and M. Gierli{\'n}ski, ``Radiative Processes, Spectral States
and Variability of Black-Hole Binaries'', Progress of Theoretical Physics
Supplement, No. 155, pp. 99-119, 2004

\bibitem{arnaud}
K. A. Arnaud, K.A., Astronomical Data Analysis Software and Systems V,
eds. Jacoby G. and Barnes J., p. 17, ASP Conf. Series vol. 101, 1996

\end{thebibliography}
\end{document}